\documentclass[%
 reprint,
superscriptaddress,
 amsmath,amssymb,
 aps,prl,
]{revtex4-2}

\usepackage{graphicx}
\usepackage{dcolumn}
\usepackage{bm}
\usepackage{siunitx}
\usepackage{mathrsfs}
\usepackage{tikz}
\usepackage{comment}
\usepackage{bbm}
\usepackage{appendix}
\usepackage{relsize}

\usepackage{soul} 

\newcommand{\bra}[1]{\left\langle#1\right|}
\newcommand{\ket}[1]{\left|#1\right\rangle}
\newcommand{\braket}[2]{\left\langle #1 \right|\left. \! #2 \vphantom{#1} \right\rangle}
\newcommand{\ave}[1]{\left\langle#1\right\rangle}

\newcommand{\tr}[1]{\operatorname{tr}\left\{#1\right\}}
\newcommand{\ketbra}[2]{\left|#1\right\rangle\! \left\langle#2\right|}

\begin{document}

\preprint{APS/123-QED}

\title{Quantum channel for modeling spin-motion dephasing in Rydberg chains}

\author{Christopher Wyenberg}
\affiliation{Institute for Quantum Computing, University of Waterloo, Waterloo, Ontario N2L 6R2, Canada.}
\affiliation{Department of Physics and Astronomy, University of Waterloo, Waterloo, Ontario N2L 6R2, Canada.}

\author{Kent Ueno}
\affiliation{Institute for Quantum Computing, University of Waterloo, Waterloo, Ontario N2L 6R2, Canada.}
\affiliation{Department of Physics and Astronomy, University of Waterloo, Waterloo, Ontario N2L 6R2, Canada.}
    
\author{Alexandre~Cooper}
\email[]{alexandre.cooper@uwaterloo.ca}
\affiliation{Institute for Quantum Computing, University of Waterloo, Waterloo, Ontario N2L 6R2, Canada.}
\affiliation{Department of Physics and Astronomy, University of Waterloo, Waterloo, Ontario N2L 6R2, Canada.}

\date{\today}

\begin{abstract}
We introduce a quantum channel to model the dissipative dynamics resulting from the coupling between spin and motional degrees of freedom in chains of neutral atoms with Rydberg interactions. The quantum channel acts on the reduced spin state obtained under the frozen gas approximation, modulating its elements with time-dependent coefficients. These coefficients can be computed exactly in the perturbative regime, enabling efficient modeling of spin-motion dephasing in systems too large for exact methods. We benchmark the accuracy of our approach against exact diagonalization for small systems, identifying its regime of validity and the onset of perturbative breakdown. We then apply the quantum channel to compute fidelity loss during transport of single-spin excitations across extended Rydberg chains in regimes intractable via exact diagonalization. By revealing the quantum-classical crossover, these results establish a bound on the maximum chain length for efficient entanglement distribution. The quantum channel significantly reduces the complexity of simulating spin dynamics coupled to motional degrees of freedom, providing a practical tool for estimating the impact of spin-motion coupling in near-term experiments with Rydberg atom arrays.
\end{abstract}

\maketitle

Neutral atoms in arrays of optical traps have emerged as a powerful platform for exploring quantum information processing~\cite{Saffman2010, Browaeys2020} including logical quantum computation~\citep{Bluvstein2024} and quantum many-body dynamics~\citep{Bernien2017, Bornet2023, Choi2023, Eckner2023}. Each atom encodes quantum information in its internal energy states, enabling single-qubit control and robust entangling operations via the Rydberg blockade mechanism~\cite{Lukin2001}. In most experiments, atoms are released from their traps before being excited to a Rydberg state due to its anti-trapping polarizability~\citep{Saffman2005}. Although ground-state wavefunctions are effectively decoupled from motional degrees of freedom, spatial gradients in the Rydberg interaction potential can lead to spin-motion coupling~\cite{Li2013, Mehaignerie2023}. Even after release from the motional ground state of a harmonic trap, the spatial wavefunction can spread significantly in free space. Because kinematic observables are typically not measured~\cite{Brown2023}, the atoms must be treated as an open quantum system with the environment provided by the continuous kinematic modes.

Accurately modeling spin-motion dephasing for Rydberg atom arrays evolving in free space is challenging due to the enormous Hilbert spaces required to describe their motional degrees of freedom. Each internal spin state must be paired with $\sim10^2$ motional modes to achieve numerical convergence, regardless of whether the basis is chosen to be the position, momentum, or harmonic oscillator modes.
This difficulty is typically bypassed by invoking the frozen gas approximation~\cite{Mourachko1998}, which neglects wavepacket dynamics during evolution.
Going beyond this approximation has proven difficult: analytical treatments have been largely restricted to small system sizes~\cite{Mehaignerie2023} or to trapped atoms~\citep{Li2013}, whereas semi-classical methods have struggled to capture late-time, highly entangled dynamics~\cite{Zhang2024}. These restrictions have so far prevented accurate modeling of spin-motion dephasing in large systems. They also have hindered the development of effective mitigation protocols, such as those based on optimal control theory~\citep{Goerz2014,Ohtsuki1999}.

\begin{figure}
\includegraphics[width=8.6cm]{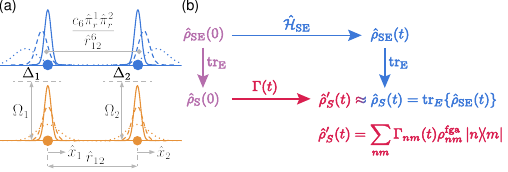}
\caption{
\label{fig:diagrams}
\textbf{Conceptual overview}.
(a)~Schematic representation of the spin-motion dynamics in a driven two-atom system with van der Waals interactions. Partial wavefunctions of atoms in the ground state (top, blue) and Rydberg state (bottom, orange) spread following release from their traps. Rydberg atoms experience spin-motion coupling due to the gradient of the interaction potential. Snapshots are shown at three time points: immediately after release (solid lines), at an intermediate time (dashed lines), and at a later time (dotted lines). The symbols are defined in the main text.
(b)~The quantum channel approximates the reduced spin state after evolution under spin-motion coupling by modulating the density matrix elements of the state obtained under the frozen gas approximation with time-dependent coefficients.
}
\end{figure}

In this work, we introduce a method to efficiently model the dissipative dynamics arising from spin-motion coupling in Rydberg systems beyond the frozen gas approximation. This method is valid in the regime of small wavefunction spread and displacement, typically realized for atoms separated by a few microns, excluding strongly interacting cases achieved with pulsed lasers~\cite{Nascimbene2015, Bharti2024}. Specifically, we introduce a quantum channel, $\Gamma(t)$, that maps the non-dephased spin state $\hat{\rho}^\text{fga}_\text{S}(t)$ onto a dephased spin state, $\hat{\rho}'_\text{S}(t) \approx \hat{\rho}_\text{S}(t)=\text{tr}_\text{E} \left[\hat{\rho}_\text{SE}(t)\right]$, which closely approximates the reduced spin state obtained after tracing out the environment (see Fig.~\ref{fig:diagrams}b). 
Its coefficients are given by
\begin{eqnarray}
    \hat{\rho}'_{n m}(t) &= \Gamma_{nm}(t) \rho_{nm}^\text{fga}(t),
\end{eqnarray}
where the indices describe the eigenstates of the unperturbed spin Hamiltonian, $\{\ket{n}\}$. The elements of the quantum channel are given by products of time-dependent coefficients, $\Gamma_{nm}(t) = C^*_n(t) C_m(t) \gamma_{nm}(t)$, which can be computed exactly with relatively low computational overhead~\cite{supplemental_materials}, scaling as $\mathcal{O}(2^L L^3)$. The channel thus offers a significant improvement over the exponential scaling of full evolution, which typically require as much as $\sim 10^2$ kinematic degrees of freedom per atom to ensure convergence.


\textit{Theory}---We consider a chain of $L$ atoms held in one-dimensional harmonic traps centered at $\ave{\hat{r}_l}_0=r^0_l$, where $\hat{r}_l$ is the position operator of the $l$th atom. We denote the relative displacement of the $l$th atom from the center of its trap as $\hat{x}_l = \hat{r}_l - r^0_l$, where $\ave{\hat{x}_l}_0=0$. The relative distance between two atoms is given by $\hat{r}_{l k} = \hat{r}_l - \hat{r}_k = r^0_{l k}+(\hat{x}_l-\hat{x}_k)$, where $r^0_{l k} = r^0_l - r^0_k$ is the relative distance between their traps. Each atom is prepared in the motional ground state of its trap, such that its initial wavepacket is $\braket{\hat{x}_l}{\psi}_0=\exp{\left[-x_l^2/2\sigma^2_0(\nu_\text{t})\right]}$, where  $\sigma_0(\nu_\text{t})=\sqrt{\hbar/2\pi\nu_\text{t}m}$ for a trap frequency $\nu_\text{t}$. In the absence of external forces, upon instantaneous release from its trap, a free wavepacket spreads as $\sigma(t) = \sigma_0 \sqrt{1 + \left(\hbar t / 2 m \sigma_0^2\right)^2}$; however, this diffusive behavior is modified by repulsive forces between Rydberg atoms and by spin-motion coupling as described below.

We consider the case where each atom is coherently driven between its ground state $\ket{g}$ and excited Rydberg state $\ket{r}$. The single-spin Hamiltonian is $\hat{\mathcal{H}}_l=\Omega_l \hat{\sigma}_x^l/2 - \Delta_l \hat{\pi}^l_r$, where $\hat{\sigma}_x^l$ is the Pauli-$x$ matrix and $\hat{\pi}^l_r=\ket{r_l}\!\bra{r_l}$ is the projection operator onto $\ket{r}$ for the $l$th atom. The interaction between two Rydberg atoms is given by $\hat{V}_{lk}\left(\hat{r}_{l k}\right) = c_\alpha \hat{r}_{l k}^{-\alpha}$, where $\alpha=3~\text{and}~6$ for dipolar and van der Waals interactions, respectively. The full Hamiltonian, including both spin and motional degrees of freedom, is
\begin{align}\label{eq:ham_se}
    \hat{\mathcal{H}}_\text{SE} &= 
    \sum_l \hat{\mathcal{H}}_l  + \sum_{l<k} V_{lk}^{(0)} \hat{\pi}^l_r \hat{\pi}^k_r  \nonumber \\
    &\quad - \sum_{l<k} F_{lk}(\hat{x}_k - \hat{x}_l) \hat{\pi}^l_r \hat{\pi}^k_r + \sum_{l} \frac{\hat{p}^2_l}{2m},
\end{align}
where $\hat{p}_l$ is the conjugate momentum operator to $\hat{x}_l$. We linearize the interaction potential near $r_{lk}^0$ to obtain the static interaction potential $V_{lk}^{(0)}=V_{lk}(r^0_{lk})$ and the linearized force $F_{lk}=-\left.{\partial V_{lk}}/{\partial r_{lk}}\right|_{r^0_{lk}}$. This approximation is valid up to first order in $\hat{x}_l-\hat{x}_k$, i.e., for wavepacket spreads much smaller than $r^0_{lk}$.

The first two terms of $\hat{\mathcal{H}}_\text{SE}$ correspond to the typical Hamiltonian under the frozen gas approximation, which can be diagonalized as $\hat{\mathcal{H}}_\text{fga} = \sum_{n} E_n \hat{\pi}_n \otimes \mathbbm{1}_{\mathbf{x}}$, where $\hat{\pi}_n = \ket{n}\!\bra{n}$ is the projector onto the bare spin eigenstate $\ket{n}$ with corresponding energy $E_n$, and $\mathbbm{1}_{\mathbf{x}} = \int \text{d}^L \mathbf{x} \ket{\mathbf{x}}\!\bra{\mathbf{x}}$ in the position representation. 
The fourth kinematic term, which commutes with $\hat{\mathcal{H}}^\text{fga}$, is responsible for the spatial diffusion of the wavepacket. Although none of these three terms couple spin and motion directly, the third linear-potential term induces position-dependent energy shifts, which we model to second order as~\cite{supplemental_materials}
\begin{equation}
    E_n^\prime(\mathbf{x}) = 
    E_n - \mathbf{f}^\intercal_n \mathbf{x} + \mathbf{x}^\intercal M^\mathbb{R}_n \mathbf{x},\label{eq:quadratic_energy}
\end{equation}
where $\mathbf{x}$ is a length-$L$ vector containing the position coordinates of all atoms. The energy shift is expressed in terms of a force vector, $\mathbf{f}^\intercal_n$, and a real-valued matrix, $M^\mathbb{R}_n$, which describes virtual transitions between spin eigenstates induced by the position-dependent potential. Their elements are given by $f^l_n = -w^l_{n n}$ and $M^{lk}_n = \sum_{m \neq n} \frac{w^l_{mn} w^k_{nm}}{E_n - E_m}$, where $w^l_{nm} = F_{l-1,l} \bra{n} \hat{\pi}^{l-1}_r \hat{\pi}^l_r \ket{m} - F_{l,l+1} \bra{n} \hat{\pi}^l_r \hat{\pi}^{l+1}_r \ket{m}$ when restricting the analysis to nearest-neighbor interactions. The diagonal force coefficients, $w^l_{nn}$, quantify the spatial asymmetry in the interaction energy of the $n$th frozen gas eigenstate with respect to the $l$th atom, whereas the off-diagonal mixing coefficients, $w^l_{n \neq m}$, quantify the spatial asymmetry in each virtual transition. These off-diagonal mixing coefficients are strictly zero when the transverse driving field is zero ($\Omega=0$); however, non-zero driving fields lead to complex spin-motion dynamics coupled across distinct frozen gas eigenstates.

The quadratic form of $E_n^\prime(\mathbf{x})$ enables us to find an elegant solution to the Schrödinger equation for the spin-motion coupled state $\ket{\Psi(t)}$ expressed in terms of partial wavefunctions, $\psi_n(\mathbf{x},t)=\braket{n,\mathbf{x}}{\Psi(t)}$. The evolution of each $\psi_n$ is governed by 
\begin{equation}
    i\hbar~{\partial_t \psi_n} = -\frac{1}{2 m}\nabla^2 \psi_n - \mathbf{f}^\intercal_n \mathbf{x} \psi_n + \mathbf{x}^\intercal M^\mathbb{R}_n \mathbf{x} \psi_n \label{eq:schr_eqns},
\end{equation}
which is a diffusion equation having linear force and quadratic potential terms. Analytical expressions for the wavefunctions can be obtained by expressing them in terms of decoupled normal modes and propagating them with the Green’s function of the quantum harmonic oscillator \cite{supplemental_materials}. The result is a collection of $\psi_n \left(\mathbf{x}, t\right)$ wavefunctions, each depending on all $x_l$ in a form that is not, in general, separable. From these expressions, we can compute the internal spin dynamics by taking the trace over kinematic degrees of freedom. These calculations involve computing overlap integrals,
\begin{eqnarray}
    \Gamma_{n m}(t) &\equiv& \int \text{d}^L x \: \psi^*_n \left(\mathbf{x},t\right) \psi_m\left(\mathbf{x},t\right)\\
    ~&=&C^*_n(t) C_m(t) \gamma_{n m}(t),
\end{eqnarray}
whose coefficients can be found exactly, but do not admit a simple expression due to the underlying algorithm being comprised by linear algebra methods (diagonalization and Gram-Schmidt orthogonalization). The $C_n(t)$ are algebraic time-dependent functions for each eigenstate $\ket{n}$. The main source of complexity lies in computing the $\gamma_{nm}(t)$ coefficients, which involve three operations (see~\citep{supplemental_materials} for details): (i)~At $t=0$, for each $\ket{n}$, we diagonalize an $(L-1)$-dimensional quadratic coupling tensor and store its diagonal elements and change of basis for later use. (ii)~For $t>0$, for each $\ket{n}$, we apply a time-dependent function to each stored diagonal element and use the stored change of basis to construct a matrix $K_n(t)$. We similarly construct a time-dependent vector $\mathbf{b}_n(t)$. (iii)~For each pair $\ket{n}, \ket{m}$, we construct the matrix $K_{n m}(t) = K^*_n(t) + K_m(t)$ and the vector $\mathbf{b}_{n m}(t)=\mathbf{b}^*_n+\mathbf{b}_m$, which are used to evaluate
\begin{equation}
    \gamma_{n m}(t) = \frac{e^{\mathbf{b}^\text{T}_{n m} K^{-1}_{n m} \mathbf{b}_{n m} / 2}}{\sqrt{\text{det} K_{n m} }}.
\end{equation}

\begin{figure}[t!]
\includegraphics[trim={0mm 0mm 0mm 0mm}, clip, width=4.27cm]{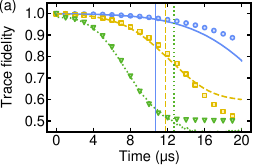}
\includegraphics[trim={0mm 0mm 0mm 0mm}, clip, width=4.27cm]{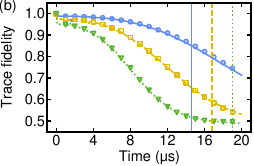} \\
\vspace{3mm}
\includegraphics[trim={0mm 0mm 0mm 0mm}, clip, width=4.27cm]{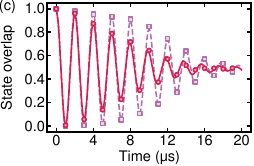}
\includegraphics[trim={0mm 0mm 0mm 0mm}, clip, width=4.27cm]{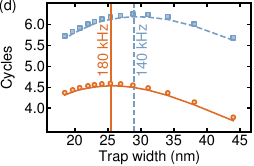}
\caption{\label{fig:2at_plots}
\textbf{Spin-motion dephasing for two atoms.}
(a)~Trace fidelity computed using the quantum channel (markers) and exact diagonalization (curves) for resonant blockaded oscillations with inter-trap spacings $r^0_{12} = \SI{2.8}{\micro\meter}$ (blue circles), $\SI{3.2}{\micro\meter}$ (yellow squares), and $\SI{3.6}{\micro\meter}$ (green triangles). Vertical lines indicate the onset of perturbative breakdown. Results are obtained for $\Delta=0$, $\Omega=\SI{10}{\MHz}$, and $\nu_\text{t}=\SI{100}{\kHz}$. 
(b)~Trace fidelity computed using the quantum channel (markers) and exact diagonalization (curves) for spin-exchange dynamics at the same inter-trap spacings as in (a). Results are obtained for $\Delta/V^{(0)}=-1/3$, $\Omega=\SI{7}{\MHz}$, and $\nu_\text{t}=\SI{100}{\kHz}$.
(c)~Time-dependent state overlap with $\ket{rg}$ for spin-exchange dynamics for $\nu_\text{t}=\SI{20}{\kHz}$ (red circles) and $\SI{200}{\kHz}$ (purple squares) with $r^0_{12} = \SI{3.2}{\micro\meter}$.
(d)~Number of spin-exchange cycles achievable with fidelity of $75\%$ (orange circles) and $50\%$ (blue squares) for different trap widths with $r^0_{12}=\SI{3.2}{\micro\meter}$. Vertical lines indicate optimal trap widths and their corresponding trap frequencies.
}
\end{figure}

\textit{Results}---To illustrate the predictions of our quantum channel and to benchmark its validity, we start with a minimal system of two atoms. This choice permits both theoretical analysis and exact diagonalization for direct comparison. We choose a global driving field $\Omega \ll V_{12}^{(0)}$ (strong blockade condition) such that, for a given global detuning $\Delta$, there exists one eigenstate $\ket{rr^\prime}$ nearly parallel to $\ket{rr}$ and three eigenstates $\ket{n_\perp}$ nearly perpendicular to $\ket{rr}$. We are interested in modeling the dynamics within the blockaded subspace $\mathcal{B}=\text{Span}\left\{\ket{n_\perp}\right\}$ after releasing the atoms from their traps.

At resonance ($\Delta = 0$), the blockaded subspace is $\mathcal{B} = \{ \ket{\pm}, \ket{a} \}$, where $\ket{\pm} = \left(c^\pm_{gg} \ket{gg} \pm c^\pm_s \ket{s}\right)/\sqrt{2} + \epsilon^\pm \ket{rr}$, with $c^\pm_{gg},~c^\pm_s \approx 1$ and $\epsilon^\pm \ll 1$. The symmetric and antisymmetric singly excited states are defined as $\ket{s} = (\ket{rg} + \ket{gr})/\sqrt{2}$ and $\ket{a} = (\ket{rg} - \ket{gr})/\sqrt{2}$, respectively. The initial state $\hat{\rho}_0 = \ket{gg}\!\bra{gg}$ undergoes blockaded oscillations between $\ket{gg}$ and $\ket{s}$; however, spin-motion coupling degrades coherence between $\ket{+}$ and $\ket{-}$ due to phase variations across motional wavefunctions, such that the system eventually reaches the mixed state $\frac{1}{2}\left(\ketbra{+}{+} + \ketbra{-}{-}\right) \approx \frac{1}{2}\left(\ketbra{gg}{gg} + \ketbra{s}{s}\right)$. We confirm this behavior by computing the trace fidelity, $F(t)=\tr{\rho(t)'\rho^\text{fga}(t)}$, between the state obtained with and without dephasing, $\rho(t)'$ and $\rho^\text{fga}(t)$ respectively, as shown in Fig.~\ref{fig:2at_plots}(a). 

The quantum channel accurately reproduces the exact dynamics at early times; however, it becomes inaccurate beyond a critical time $T^*_{n_\perp}$ due to a breakdown of the perturbative regime. We estimate $T^*_{n_\perp}$ by approximating the wavefunction displacement and its effect on internal energy perturbations. In the case of two atoms, the third, spin-motion coupling term of Eq.~\eqref{eq:ham_se} reduces to $F_{12} \hat{x}_{12} \hat{\pi}^1_r \hat{\pi}^2_r$, where $x_{12}=x_2-x_1$ is the single independent degree of freedom~\cite{supplemental_materials}. The resulting perturbation to the bare energies $E_n$ involves terms $\hat{x}_{12} w_{mn} = \bra{m} \hat{x}_{12} F_{12} \hat{\pi}^1_r \hat{\pi}^2_r \ket{n}$. Perturbation theory is only valid as long as $\left|\hat{x}_{12} w_{mn}\right| < \left|E_n - E_m\right|$~\cite{Cohen-Tannoudji:1977}. To estimate $\left|\hat{x}_{12} w_{mn}\right|$, we consider the mean position of the partial wavepacket $\psi_m(x_{12},t)$. Exact numerical calculations confirm that the mean separation $\bar{x}_{rr'}(t) = \int x_{12}|\psi_{rr'}|^2 \textrm{d}x_{12}$ undergoes a large displacement $\bar{x}_{rr'}(t) \approx F_{12} t^2/2 m$, consistent with Ehrenfest's theorem~\citep{Cohen-Tannoudji:1977}. This displacement arises from the large repulsive force $f_{rr'}=F_{12} \bra{rr'} \hat{\pi}^1_r \hat{\pi}^2_r \ket{rr'} \approx F_{12}$ present in Eq.~\eqref{eq:schr_eqns}. Perturbation theory no longer describes the contribution of $\ket{rr'}$ to $E_{n_\perp}$ beyond the time $T^*_{n_\perp}$ at which $\bar{x}_{rr'}(T^*_{n_\perp}) F_{12} \left|\bra{n_\perp} \hat{\pi}^1_r \hat{\pi}^2_r \ket{rr'}\right| = \left|E_{rr'}-E_{n_\perp}\right|$. For van der Waals interactions ($\alpha=6$), we obtain
\begin{equation}
    T^*_{n_\perp} = \frac{r^0_{12}}{3V^{(0)}}\sqrt{\frac{m}{2} \left|\frac{E_{rr'}-E_{n_\perp}}{\braket{n_\perp}{rr'}}\right|}.
\end{equation}
To indicate the onset of perturbative breakdown, we display $T^* = \min_{n_\perp}(T^*_{n_\perp})$ as vertical lines in Fig.~\ref{fig:2at_plots}~(a,b). 

Far from resonance ($\left|\Delta\right| \gg \Omega$), 
the spin Hamiltonian maps onto an effective Heisenberg~XX model, $\hat{\mathcal{H}}_\text{eff} = J(\hat{\sigma}^+_1 \hat{\sigma}^-_2 + \hat{\sigma}^-_1 \hat{\sigma}^+_2)$, where the synthetic spin-exchange rate is given by $J=\Omega^2 V^{(0)}/4\Delta(\Delta-V^{(0)})$~\citep{Yang2019, Kim2024, Ueno2025}. This Hamiltonian drives coherent oscillations between $\ket{rg}$ and $\ket{gr}$ that are damped by spin-motion dephasing (Fig.~\ref{fig:2at_plots}(b)). We observe that the dephasing rate increases with interatomic distance, because weaker interactions at larger separations lead to greater leakage out of the single-spin excitation subspace, which in turn increases the sensitivity of the perturbed eigenstates to position-dependent phases. We also observe that the dephasing rate decreases with trap depth (see Fig.~\ref{fig:2at_plots}(c)), because tighter confinement reduces the spatial extent of the relative coordinate wavepacket, thereby suppressing phase variations across it.

The spin-exchange fidelity highlights a trade-off between evolution time and trap depth. Shallower traps are favorable for longer evolutions, as their narrower momentum distributions reduce spatial spreading at later times. In contrast, deeper traps are favorable for shorter evolutions, as their stronger confinement limits the spatial extent of the wavepacket, reducing early-time dephasing. As a result, there exists an optimal trap depth that maximizes the number of coherent exchange cycles achievable within a given fidelity threshold. We compute the number of exchange cycles achievable within $75\%$ and $50\%$ fidelity as a function of the initial trap width (see Fig.~\ref{fig:2at_plots}(d)). As the fidelity threshold is lowered, the optimal trap depth shifts toward shallower values.

\begin{figure}
\includegraphics[trim={0mm 0mm 0mm 0mm}, clip, width=4.27cm]{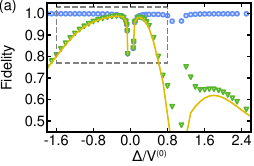}
\includegraphics[trim={0mm 0mm 0mm 0mm}, clip, width=4.27cm]{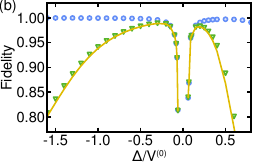} \\
\vspace{3mm}
\includegraphics[trim={0mm 0mm 0mm 0mm}, clip, width=4.27cm]{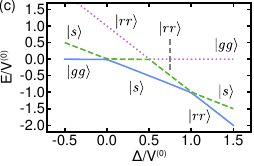}
\includegraphics[trim={0mm 0mm 0mm 0mm}, clip, width=4.27cm]{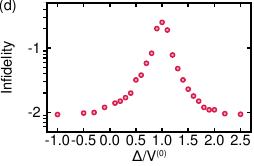}
\caption{\label{fig:2at_sweep}
\textbf{Dependence on detuning.} 
(a-b)~Spin-exchange fidelity after a single spin-exchange cycle computed using the frozen gas approximation (blue circles), the dephasing channel (green triangles), and exact diagonalization (yellow line).  
(b) Zoomed-in view on the red-detuned region indicated by the dashed box in (a).
(c)~Spectral eigenstates, excluding $\ket{a}$, labeled by their closest parallel states in the set $\left\{\ket{gg}, \ket{s}, \ket{rr}\right\}$. The performance of the quantum channel degrades for initial states nearly degenerate with the interacting $\ket{rr}$ state. 
(d)~Log infidelity of the quantum channel showing the breakdown of the perturbative approximation at $\Delta/V^{(0)}=1$ due to the degeneracy of the eigenstate nearly parallel to $\ket{s}$ and the one nearly parallel to $\ket{rr}$. All data are computed using $\Omega=\SI{10}{\MHz}$ and $r_{12}^0=\SI{2.6}{\micro\meter}$.
}
\end{figure}

To further understand the effect of spin-motion dephasing on spin-exchange dynamics, and to explore the regime of validity of the quantum channel, we compute the spin-exchange fidelity as the probability of recovering the state $\ket{rg}$ after one exchange cycle $\ket{rg} \rightarrow \ket{gr} \rightarrow \ket{rg}$ as a function of $\Delta$ (see Fig.~\ref{fig:2at_sweep}(a)). At large detuning ($|\Delta|/V^{(0)}\gg 1$), the small exchange rate $J\sim\Delta^{-2}$ leads to a long exchange period, $t_{2\pi}\sim\Delta^2$, and thus a large amount of spin-motion dephasing. At moderate detuning ($|\Delta|/V^{(0)}\lesssim1$), there exists a competition between spin-motion dephasing (reduced by faster exchanges at smaller $|\Delta|$) and the accuracy of the synthetic exchange Hamiltonian (improved at larger $|\Delta|$). An optimal detuning exists near $\Delta/V^{(0)} \approx \pm 0.25$ (see Fig.~\ref{fig:2at_sweep}(b)). Comparing the spin-exchange fidelity at blue (positive) and red (negative) detuning for similar spin-exchange magnitudes $|J|$, we find that the blue-detuned case is more sensitive to spin-motion dephasing than the red-detuned case, as it is predominantly mediated via $\ket{rr}$ rather than $\ket{gg}$.
For example, although $|J|$ is similar at $\Delta/V^{(0)} = -0.21$ and at $\Delta/V^{(0)} = +0.5$, the spin-exchange fidelity is significantly lower in the blue-detuned case (see Fig.~\ref{fig:2at_sweep}(b)).

The breakdown of the quantum channel in the blue-detuned region (see the discrepancy between the green triangles and yellow curve in Fig.~\ref{fig:2at_sweep}(a)) can be explained by examining the changes in the eigenstates with the detuning. In Fig.~\ref{fig:2at_sweep}(c), we label each spectral branch according to its closest parallel state among the states $\left\{\ket{gg}, \ket{s}, \ket{rr}\right\}$. Near $\Delta/V^{(0)}=1$, the eigenstate nearly parallel to the symmetric state $\ket{s}=(\ket{rg}+\ket{gr})/\sqrt{2}$ exhibits an avoided crossing with the state nearly parallel to $\ket{rr}$. The validity of the perturbative construction of Eq.~\eqref{eq:quadratic_energy} requires that $\left|\hat{x}_{12} F_{12} \bra{n}\hat{\pi}^1_r \hat{\pi}^2_r \ket{m}\right| < \left|E_n - E_m\right|$, a condition that is violated for the eigenstates nearly parallel to $\ket{s}$ and to $\ket{rr}$---the former state having a significant projection onto the initial state $\ket{rg}=(\ket{s}+\ket{a})/\sqrt{2}$.

The accuracy of the quantum channel over a given evolution time improves as $\Delta$ deviates from $V^{(0)}$; however, this trend is difficult to discern from Fig.~\ref{fig:2at_sweep}(a,b) due to the variation in exchange rate with $\Delta$.
Thus, to better illustrate the channel performance as a function of $\Delta$, we compute the trace distance between exact and channel results, evaluated for each $\Delta$ at the time corresponding to a fidelity loss of $20\%$. This infidelity metric is plotted in Fig.~\ref{fig:2at_sweep}(d) as a function of $\Delta$ for an initial state $\ket{rg} = (\ket{s} + \ket{a})/\sqrt{2}$. The performance degradation near $\Delta/V^{(0)}=1$ is consistent with the avoided crossing observed in Fig.~\ref{fig:2at_sweep}(c).

\textit{Quantum-classical crossover}---To demonstrate the applicability of our method for modeling dissipative dynamics in otherwise intractable regimes, we use the quantum channel to predict the quantum-to-classical crossover when distributing entanglement over Rydberg chains containing tens of atoms~\cite{Yang2019, Ueno2025}. For large detuning, $|\Delta|\gg\Omega$, the Rydberg chain supports spin transport in the single-spin excitation subspace via the effective Hamiltonian $\hat{\mathcal{H}}_\text{eff} = \sum_l \mu_l \hat{\sigma}^+_l \hat{\sigma}^-_l + \sum_{l<k} J_{l k}\left(\hat{\sigma}^+_l \hat{\sigma}^-_k + \hat{\sigma}^+_k \hat{\sigma}^-_l\right)$, where $\mu_l$ and $J_{lk}$ depend on the set of optimized control parameters, $\{\Omega_l, \Delta_l, r_{lk}\}$. Setting $\hat{\mathcal{H}}^\text{fga}_\text{eff} = \hat{\mathcal{H}}_\text{eff}\left(r^0_{lk}\right)$ and $\hat{\mathcal{H}}^{(1)}_{\text{eff},{lk}} = \left.\partial \hat{\mathcal{H}}_\text{eff} /\partial r_{lk}\right|_{r^0_{lk}}$, we can express the effective spin-exchange Hamiltonian as
\begin{equation}
    \hat{\mathcal{H}}^\text{SE}_\text{eff} = \hat{\mathcal{H}}^\text{fga}_\text{eff}
    + \sum_{l k} \hat{\mathcal{H}}^{(1)}_{\text{eff},{lk}} \left(\hat{x}_k-\hat{x}_l\right)
    + \sum_l \frac{\hat{p}^2_l}{2m}.\label{eq:spexch_ch_form}
\end{equation}
This Hamiltonian resembles Eq.~\eqref{eq:ham_se}, allowing its dynamics to be modeled using the quantum channel.

\begin{figure}
\includegraphics[trim={0mm 0mm 0mm 0mm}, clip, width=4.27cm]{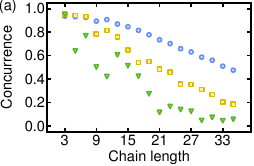} \includegraphics[trim={0mm 0mm 0mm 0mm}, clip, width=4.27cm]{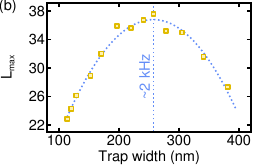}
\caption{\label{fig:varlen_plots}
\textbf{Entanglement transport.} 
(a)~Concurrence between the $0$th auxiliary spin and the $L$th spin computed using the quantum channel for trap frequencies $\nu_\text{t}=\SI{3}{\kHz}$ (blue circles), $\SI{10}{\kHz}$ (yellow squares), and $\SI{30}{\kHz}$ (green triangles).
(b)~Quantum-classical crossover, defined by the maximum chain length $L_\text{max}$, computed as a function of trap width using the quantum channel (yellow squares) with quadratic fit (dotted curve). The vertical line denotes the maximum of the quadratic fit, annotated with its corresponding trap frequency. Data are obtained for $r^0_{12} = \SI{3.2}{\micro\meter}$, $\Delta/V^{(0)}=-1/3$, and $\Omega=|\Delta| / 10$.
}
\end{figure}

We consider the specific problem of coherently distributing an entangled Bell pair across a Rydberg spin chain. For a given chain of $L$ equidistant spins, there exists a set of control parameters, $\{\Omega_l, \Delta_l\}$, that satisfies the perfect transport condition, enabling the transfer of a single-spin excitation from one end of the chain to the other~\citep{Yang2019, Ueno2025}. We initialize the system in the state $(\ket{r_0 g_1}+\ket{g_0 r_1}) \otimes \ket{g_2 g_3 \dots g_L} / \sqrt{2}$ and let the chain transport the state of the first spin to the $L$th spin. The entanglement between the auxiliary and end spins can be detected using an entanglement witness~\cite{Peres1996, Horodecki1996, Sun2020}, as well as quantified using an entanglement measure, which we choose as the concurrence~\cite{Hill1997, Wootters1998, Hildebrand2007, Horodecki2009}, $C_{0L}(t) = C\left[\hat{\rho}_{0L}(t)\right]$, where $\hat{\rho}_{0L}(t)$ is the reduced density matrix for sites $0, L$ at time $t$. 

We compute the maximum concurrence $C_{0L}$ achieved as a function of chain length for atoms initially confined in traps of varying depth (see Fig.~\ref{fig:varlen_plots}(b)). Under the unitary evolution generated by $\hat{\mathcal{H}}^\text{fga}_\text{eff}$, there exists a time $t_\pi$ at which $C_{0L}(t_{\pi})=1$; however, in the presence of spin-motion coupling, there exists a maximum chain length beyond which $C_{0L}$ never exceeds $1/2$. For such lengths, the reduced state $\hat{\rho}_{0L}$ is predominantly a mixed state with reduced quantum coherence. We define this quantum-classical crossover as the length $L_\text{max}$ for which $C_{0L_\text{max}}=1/2$ at its peak time. We note that, while distillable entanglement may still be present beyond this point, we adopt $C_{0L} = 1/2$ as a practical threshold to define the crossover. The concurrence is maximized by a relatively shallow trap of frequency $\nu_\text{t}\sim\!\SI{2}{\kHz}$ (see Fig.~\ref{fig:2at_plots}(d)), as can be achieved, for example, by adiabatic ramp-down. This value is to be contrasted with the $\SI{140}{\kHz}$ trap frequency obtained when maximizing the $50\%$ trace fidelity for a two-atom system with similar control parameters (see Fig.~\ref{fig:2at_plots}(d)). Our results indicate that quantum coherence is more robust to motional dephasing during entanglement transport over a spin chain than during repeated spin exchange between a pair of atoms. This enhanced robustness arises because bulk atoms in the chain experience approximately equal and opposite forces from their neighbors, reducing the net motional dephasing during transport.

\textit{Conclusion}---We constructed a quantum channel to compute the dephasing in internal spin dynamics induced by spin-motion coupling. We showed that the channel accurately predicts the exact dynamics on small systems, with discrepancies explained by the breakdown of the perturbative approximation. We applied the channel to quantify spin-motion dephasing during entanglement transport in extended spin chains, identifying an optimal trap depth for maximizing the quantum-classical crossover in realistic near-term experiments. These results indicate that entanglement transport in an extended spin chain is less sensitive to spin-motion coupling than exchange dynamics of similar duration between two spins. The channel reduces numerical complexity by avoiding the need to simulate the full kinematic Hilbert space associated with spin-motion coupling.
However, the channel has three key limitations: (i) it applies only to time-independent control fields, (ii) it is restricted to valid perturbation regimes, and (iii) it does not employ advanced numerical evolution methods within the internal spin space. All of these limitations are directions for future research.

\textit{Acknowledgments}---This research was supported in part by the Canada First Research Excellence Fund (CFREF). We acknowledge funding from the Natural Sciences and Engineering Research Council of Canada (NSERC). We thank Dr. Abhijit Chakraborty for insightful discussions.


\providecommand{\noopsort}[1]{}\providecommand{\singleletter}[1]{#1}%

\section{Supplemental Material}

\subsection{Detailed channel derivation}

We detail here the dephasing channel derivation as well as its approximations. Our starting point is the first-order Taylor expansion of the Hamiltonian given in the main text as Eq.~\eqref{eq:ham_se}; for reference,
\begin{align}\label{eq:sup:ham_se}
    \widehat{\mathcal{H}}^\text{SE} &= 
    \sum_l \widehat{\mathcal{H}}^\text{ctrl}_l  + \sum_{l<k} V_{lk}^{(0)} \hat{\pi}^l_r \hat{\pi}^k_r  \nonumber \\
    &\quad - \sum_{l<k} F_{lk}(\hat{x}_k - \hat{x}_l) \hat{\pi}^l_r \hat{\pi}^k_r + \sum_{l} \frac{\hat{p}^2_l}{2m}.
\end{align}
We seek to propagate the total system time evolution generated by Eq.~\eqref{eq:sup:ham_se} to some later time $t$, in a form permitting an analytical trace over motional degrees of freedom while remaining accurate in its description of spin-motion dephasing.

We partition $\widehat{\mathcal{H}}^\text{SE}$ into its ideal frozen gas part $\widehat{\mathcal{H}}^\text{fga}_\text{S}$ which acts only in the spin Hilbert space $\mathbb{H}_\text{S}$, its diffusion part $\widehat{\mathcal{H}}^\text{diff}_\text{E}$ which acts only in the motion environment Hilbert space $\mathbb{H}_\text{E}$, and its mixing part $\widehat{\mathcal{H}}^\text{mix}_\text{SE}$ which causes mixing between the spins and the environment; respectively,
\begin{align}
    \widehat{\mathcal{H}}^\text{fga}_\text{S} &= \sum_l \widehat{\mathcal{H}}^\text{ctrl}_l  + \sum_{l<k} V_{lk}^{(0)} \hat{\pi}^l_r \hat{\pi}^k_r, \\
    \widehat{\mathcal{H}}^\text{diff}_\text{E} &= \sum_l \frac{\widehat{p}^2_l}{2m}, \text{ and} \\
    \widehat{\mathcal{H}}^\text{mix}_\text{SE} &= \sum_{l<k} V_{lk}^{(1)}(\hat{x}_l- \hat{x}_k) \hat{\pi}^l_r \hat{\pi}^k_r.
\end{align}

\textit{Effective mixing approximation}---Our first approximation is to replace the sum of $\widehat{\mathcal{H}}^\text{fga}_\text{S}$ and $\widehat{\mathcal{H}}^\text{mix}_\text{SE}$ with an effective Hamiltonian $\widehat{\mathcal{H}}^\text{eff}_\text{SE}$ via time-independent perturbation theory. This process involves the perturbation of energies and of eigenstates. Below, we perturb the energies to second order, but we neglect the perturbation of eigenstates. Exact simulation finds that eigenstate perturbation has a negligible effect upon dephasing.

\textit{Perturbation of energies}---We consider a state $\ket{n;\mathbf{x}}=\ket{n}\ket{x_1}\cdots\ket{x_L}$ which is an eigenstate of $\widehat{\mathcal{H}}^\text{fga}_\text{S}$ and perturb its energy to second order in $\widehat{\mathcal{H}}^\text{mix}_\text{SE}$; i.e.,
\begin{equation}
    E_n \rightarrow E^\prime_{n;\mathbf{x}} = E_n^{(0)} + E^{(1)}_{n;\mathbf{x}} + E^{(2)}_{n;\mathbf{x}}\label{eq:sup_pert_energy}
\end{equation}
where
\begin{align}
    E^{(0)}_n &= E_n, \\
    E^{(1)}_{n;\mathbf{x}} &= \bra{n;\mathbf{x}} \widehat{\mathcal{H}}^\text{mix}_\text{SE}
    \ket{n;\mathbf{x}}, \text{ and} \\
    E^{(2)}_{n;\mathbf{x}} &= \sum_{m \neq n} \frac{\left|\bra{m;\mathbf{x}} \widehat{\mathcal{H}}^\text{mix}_\text{SE}
    \ket{n;\mathbf{x}}\right|^2} {\left(E_n-E_m\right)}.
\end{align}
Note that, formally, the second-order term is generated by a sum which also extends over all $\ket{\mathbf{x}^\prime}$ position basis states \citep{Cohen-Tannoudji:1977}; however, the operator $\widehat{\mathcal{H}}^\text{mix}_\text{SE}$ is diagonal in the $\ket{\mathbf{x}^\prime}$ basis such that the sum reduces to the particular basis state $\ket{\mathbf{x}}$ being perturbed. The effective Hamiltonian is thus
\begin{equation}
    \widehat{\mathcal{H}}^\text{eff}_\text{SE} = \sum_n \int \text{d}^L \mathbf{x} \ket{n;\mathbf{x}}\bra{n;\mathbf{x}}
    \left(
    E_n + E^{(1)}_{n;\mathbf{x}} + E^{(2)}_{n;\mathbf{x}}
    \right).\label{eq:sup:ham_eff}
\end{equation}

\textit{Schrödinger evolution}---Our objective is to now propagate an initial state
\begin{equation}
    \ket{\Psi(0)} = \ket{\phi_0; \mathbf{G}_{\sigma_0}},
\end{equation}
having some internal spin state $\ket{\phi_0}$ and a motional state $\ket{\mathbf{G}_{\sigma_0}}$, to a later state
\begin{equation}
    \ket{\Psi(t)} = e^{-i \left(\widehat{\mathcal{H}}^\text{eff}_\text{SE} + \widehat{\mathcal{H}}^\text{diff}_\text{E}\right) t}\ket{\phi_0; \mathbf{G}_{\sigma_0}}.\label{eq:prop_approx}
\end{equation}
The evolution generated by Eq.~\eqref{eq:prop_approx} is described by a Schrödinger equation for each partial wavefunction $\Psi_n(\mathbf{x},t) = \braket{n;\mathbf{x}}{\Psi(t)}$,
\begin{align}
    i\frac{\partial \Psi_n}{\partial t} = -\frac{1}{2}\sum_{l=1}^L \frac{\partial^2 \Psi_n}{\partial x_l^2} + \left(
    E_n + E^{(1)}_{n;\mathbf{x}} + E^{(2)}_{n;\mathbf{x}}
    \right)\Psi_n \label{eq:sup_approx_schr_evo}
\end{align}
in normalized units for which $\sigma_0, \hbar^2/m\rightarrow1$. Each $\Psi_n(\mathbf{x},t)$ is not, in general, separable across distinct motional degrees of freedom $x_l$. The $E_n \Psi_n$ term generates ideal frozen gas evolution over $t$ (i.e., unmodified by motional degrees of freedom). For legibility, we temporarily suppress the overall phase evolution generated by this term.

The terms linear and quadratic in $\mathbf{x}$ are responsible for spin-motion dephasing. The linear term maintains separable state evolution across atoms, while the quadratic term mixes distinct atomic coordinates. We seek to remove this mixing by a change of coordinates. Let $E^{(2)}_{n;\mathbf{x}} = \mathbf{x}^\text{T} M_n \mathbf{x}$, such that
\begin{equation}
    M^{lk}_n = \sum_{m \neq n} \frac{w^l_{mn} w^k_{nm}}{E_n - E_m}
\end{equation}
where
\begin{equation}
    w^l_{nm} = \bra{n} \hat{\pi}^l_r \left(V^{(1)}_{l,l-1} \hat{\pi}^{l-1}_r - V^{(1)}_{l,l+1} \hat{\pi}^{l+1}_r \right) \ket{m}.
\end{equation}
Although the rank-2 tensor $M_n$ is not in general real, the quadratic coupling is independent of its imaginary parts; that is, $\mathbf{x}^\text{T} M_n \mathbf{x}=\mathbf{x}^\text{T} \text{Re}\left(M_n\right) \mathbf{x}$ since $w^l_{n m} = w^{l\:*}_{m n}$. We are thus justified in restricting our following analysis to $\text{Re}\left(M_n\right) \equiv M^\mathbb{R}_n$. In lab coordinates $\mathbf{x}$ the matrix $M^\mathbb{R}_n$ entangles distinct atomic coordinates. As $M^\mathbb{R}_n$ is real and symmetric, this coordinate entangling may be removed by diagonalizing $M^\mathbb{R}_n$.

Prior to its diagonalization, however, we observe that $M^\mathbb{R}_n$ possesses rank at most $L-1$ due to the translation symmetry $x_l \rightarrow x_l + c$ of the original Hamiltonian. There thus exists an independent degree of freedom which does not contribute to the dephasing process; namely, the sum of coordinates $\Sigma = \sum_l x_l$. We remove $\Sigma$ from the problem by performing Gram-Schmidt orthogonalization along the sequence of adjacent atomic separations. The details are provided in Section \textbf{Gram-Schmidt and disentangling transformations}, where we show that
\begin{equation}
    \mathbf{x}^\text{T} M^\mathbb{R}_n \mathbf{x} = \mathbf{g}^\text{T} S_n \mathbf{g}
\end{equation} for $S_n \in \mathbb{R}^{L-1}{\times}\mathbb{R}^{L-1}$ the quadratic coupling tensor in the Gram-Schmidt coordinates (less $\Sigma$) $\mathbf{g} \in \mathbb{R}^{L-1}$.

For each spin eigenstate $\ket{n}$, we diagonalize $S_n = Q^\text{T}_n D_n Q_n$ and denote by $d^l_n$ the $L-1$ real eigenvalues of $S_n$ appearing on the main diagonal of $D_n$. We also define the $\ket{n}$-specific coordinates $\mathbf{s}_n = Q_n \mathbf{g} \in \mathbb{R}^{L-1}$. Composing the Gram-Schmidt transformation and the $\ket{n}$-specific diagonalization, we have that $\mathbf{s}_n = T_n \mathbf{x}$ for $T_n = Q_n G \mathbf{x}$ with $G$ the Gram-Schmidt transformation constructed in Section \textbf{Gram-Schmidt and Disentangling Transformations}. The quadratic coupling is now diagonal in $\mathbf{s}_n$; i.e.,
\begin{equation}
    \mathbf{x}^\text{T} M^\mathbb{R}_n \mathbf{x} = \mathbf{s}^\text{T}_n D_n \mathbf{s}_n.
\end{equation}

As $T_n$ is orthogonal, the Laplacian is invariant under $\mathbf{x} \rightarrow \mathbf{s}_n$ such that Eq.~\eqref{eq:approx_schr_evo} becomes (recall our momentary suppression of ideal evolution)
\begin{align}
    i\frac{\partial \Psi_n}{\partial t} = -\frac{1}{2}\sum_{l=1}^{L-1} \frac{\partial^2 \Psi_n}{\partial s^2_{n,l}} - \mathbf{F}^\text{T}_n \mathbf{s}_n \Psi_n + \mathbf{s}^\text{T}_n D_n \mathbf{s}_n \Psi_n \label{eq:approx_schr_evo}
\end{align}
where $\mathbf{F}_n$ is the force-like term corresponding to the transformation of $E^{(1)}_{n;\mathbf{x}}$ to $\mathbf{s}_n$ coordinates,
\begin{equation}
    F^l_n = -\sum_k T^{lk}_n w^k_{nn}.
\end{equation}

\textit{Solution}---As the initial motional wavefunction of the system at $t=0$ is a separable product of Gaussians in $\mathbf{x}$, and as a separable product of Gaussians is invariant under an orthogonal transformation, the system commences in a separable product of Gaussians in $\mathbf{s}_n$. Furthermore, Eq.~\eqref{eq:approx_schr_evo} does not couple distinct coordinates $s_{n,l}$, so that our model generates separable evolution in $\mathbf{s}_n$ for all $t>0$. We may therefore describe the wavefunction as a product
\begin{equation}
    \Psi_n\left(\mathbf{s}_n,t\right) = \psi^1_n(s_{n,1},t) \psi^2_n(s_{n,2},t)\dots \psi^{L-1}_n(s_{n,L-1},t),
\end{equation}
whose each $\psi^l_n(s_l,t)$ evolves according to
\begin{equation}
    i \frac{\partial \psi^l_n}{\partial t} = -\frac{1}{2} \frac{\partial^2 \psi^l_n}{\partial s^2_{n,l}} - F^l_n s_{n,l} \psi^l_n + \frac{\omega^{l\:2}_n}{2} s^2_{n,l} \psi^l_n
\end{equation}
for $\omega^l_n = \sqrt{d^l_n}$.

Let us change to the natural coordinates
\begin{equation}
z_{n,l} = \sqrt{\omega^l_n} s_{n,l} - F^l_n / {\omega^l_n}^{\frac{3}{2}}
\text{ and }
\tau^l_n = \omega^l_n t
\end{equation}
such that
\begin{equation}
    i\frac{\partial \psi^l_n}{\partial \tau^l_n} = -\frac{1}{2} \frac{\partial^2 \psi^l_n}{\partial z^2_{n,l}} + \frac{1}{2} z^2_{n,l} \psi^l_n - \frac{F^{l\:2}_n}{2\omega^{l\:3}_n} \psi^l_n. \label{eq:qho_ham}
\end{equation}
The rightmost term is a shift in energy which may be factored off by transforming to a rotating frame; i.e., by letting
\begin{equation}
    \psi^l_a(z_{n,l},\tau) = e^{i F^{l\:2}_n \tau^l_n / 2\omega^{l\:3}_n} \tilde{\psi}^l_n(z_{n,l},\tau^l_n).
\end{equation}
The remaining part of Eq.~\eqref{eq:qho_ham} is the Hamiltonian of a quantum harmonic oscillator in natural units, which we propagate exactly using the method of Green's functions. We thus have, in the rotating frame,
\begin{equation}
    \tilde{\psi}^l_n(z_{n,l},\tau^l_n) = \int \text{d} \bar{z}_{n,l} K_\text{H}(z_{n,l},\bar{z}_{n,l};\tau^l_n) \psi^l_n(\bar{z}_{n,l},0) \label{eq:green_integral}
\end{equation}
for
\begin{equation}
    \psi^l_n(z_{n,l},0) = \frac{1}{\sqrt[4]{\pi}} e^{-(z_{n,l} + F^l_n/\omega^{l\:3/2}_n)^2/ 2\omega^l_n},
\end{equation}
where
\begin{equation}
    K_\text{H}(z,\bar{z};\tau) = \frac{1}{\sqrt{2\pi i \sin \tau}} e^{\frac{i}{2} \left[\left(z^2 + \bar{z}^2\right) \cot \tau - \frac{2 z \bar{z}}{\sin \tau}\right]}
\end{equation}
is the Green's function of the quantum harmonic oscillator, conventionally referred to as the Feynman propagator.

Upon integrating Eq.~\eqref{eq:green_integral} and re-introducing the rotating frame factor we obtain
\begin{equation}
    \psi^l_n\left(s_{n,l},t\right) = \frac{1}{\sqrt[4]{\pi}} \frac{e^{\alpha^l_n}}{\sqrt{\gamma^l_n}} e^{-s_{n,l} \bar{K}^{l l}_n s_{n,l} / 2 + \bar{b}^l_n s_{n,l}}, \label{eq:sup_psi_final}
\end{equation}
where
\begin{align}
    \bar{K}^{l l}_n &= \frac{i \omega^l_n}{\gamma^l_n \sin \omega^l_n t} - i \omega^l_n \cot \omega^l_n t, \label{eq:sup_kbar} \\
    \bar{b}^l_n &= i\frac{F^l_n}{\omega^l_n}\left[
    \frac{1 + i \left(\sin \omega^l_n t\right) / \omega^l_n}{\gamma^l_n \sin \omega^l_n t}
    - \cot \omega^l_n t
    \right], \\
    \alpha^l_n &= \frac{i}{2}\frac{{F^l_n}^2}{{\omega^l_n}^3} \left\{ \vphantom{\frac{{\left[\left(\right)\right]}^2}{B}} \omega^l_n t + \cot\omega^l_n t + \frac{i}{\omega^l_n} \right. \nonumber \\
    &\hspace{15mm} \left. - \frac{\left[1+i\left(\sin \omega^l_n t\right) / \omega^l_n\right]^2}{\gamma^l_n \sin \omega^l_n t} \right\} \text{, and} \\
    \gamma^l_n &= \cos \omega^l_n t + i \left(\sin \omega^l_n t\right) / \omega^l_n. \label{eq:sup_gamma}
\end{align}
The above expressions become singular at $\omega^l_n=0$; however, the limiting behaviour is well defined, as the singularities turn out to be removable as detailed in Section \textbf{Removable singularities}.

The composite motional wavefunction across all $\mathbf{g}$ coordinates is now
\begin{equation}
    \psi_n \left(\mathbf{g}, t\right) = \frac{1}{\pi^{N/4}} C_n e^{-\mathbf{g}^\text{T} K_n \mathbf{g} / 2 + \mathbf{b}^\text{T}_n \mathbf{g}},
\end{equation}
where
\begin{align}
    C_n &= \prod_l \frac{e^{\alpha^l_n}}{\sqrt{\gamma^l_n}}, \\
    K_n &= Q^\text{T}_n \bar{K}_n Q_n \text{, and} \\
    \mathbf{b}_n &= Q^\text{T}_n \bar{\mathbf{b}}_n.
\end{align}
The $\bar{K}_n$ matrix is the diagonal matrix having elements $\bar{K}^{l k}_n = \delta^{lk} \bar{K}^{ll}_a$.

The trace over motional degrees of freedom yields a dephasing factor for the reduced spin state density matrix element $\rho^\prime_{nm}$ given by the overlap integral
\begin{equation}
    \Gamma_{nm} = \frac{1}{\pi^{N/2}} C^*_n C_m \int \text{d}^{L-1} g  e^{-\mathbf{g}^\text{T} K_{nm} \mathbf{g} / 2 + \mathbf{b}_{nm}^\text{T} \mathbf{g}}, \label{eq:overlap}
\end{equation}
where
\begin{equation}
    K_{nm} = K^*_n + K_m \text{ and } \mathbf{b}_{nm} = \mathbf{b}^*_n + \mathbf{b}_m.
\end{equation}
We prove in Section \textit{Proof that $K_{ab}$ is invertible} that $K_{ab}$ is invertible; as it is symmetric, $\left[K^{-1}_{nm}\right]^\text{T} = K^{-1}_{nm}$ and the integral may be written
\begin{align}
    \Gamma_{nm} &= \frac{1}{\pi^{N/2}} C^*_n C_m e^{\mathbf{b}^\text{T}_{nm} K^{-1}_{nm} \mathbf{b}_{nm}/2} \nonumber \\
    &\quad \times \int \text{d}^{L-1} g e^{-\left(\mathbf{g}^\text{T} + \mathbf{b}^\text{T}_{nm} K^{-1}_{nm}\right) K_{nm} \left(\mathbf{g} + K^{-1}_{nm} \mathbf{b}_{nm}\right)/2}.
\end{align}
After a change of variables $\mathbf{g}\rightarrow \mathbf{g}+K^{-1}_{nm} \mathbf{b}_{nm}$ the integral evaluates to
\begin{equation}
    \Gamma_{nm} = 2^\frac{L-1}{2} C^*_n C_m \frac{e^{\mathbf{b}^\text{T}_{nm} K^{-1}_{nm} \mathbf{b}_{nm} / 2}}{\sqrt{\text{det} K_{nm} }}.
\end{equation}

\subsection{Removable singularities}

Eqs. \eqref{eq:sup_kbar}–\eqref{eq:sup_gamma} possess singularities at $\omega^l_n=0$; however, each singularity is removable. The singularity in $\gamma^l_n$ is easily removed; by inspection,
\begin{equation}
    \lim_{\omega^l_n \rightarrow 0} \gamma^l_n(\omega^l_n) = 1 + i t.
\end{equation}
The singularities in $\bar{K}^{ll}_n, \bar{b}^l_n,$ and $\alpha^l_n$ are removed by evaluating their Laurent series expansion with respect to $\omega^l_n$. Indeed, upon expanding we find that all negative order terms vanish, and that the zeroth order terms give 
\begin{align}
    \lim_{\omega^l_n \rightarrow 0} \bar{K}^{ll}_n &= \frac{1}{1+it}, \label{eq:sup_lim_bark} \\
    \lim_{\omega^l_n \rightarrow 0} \bar{b}^l_n &= \frac{F^l_n \left(2i - t\right) t}{2\left(1+it\right)}, \text{ and} \label{eq:sup_lim_barb} \\
    \lim_{\omega^l_n \rightarrow 0} \alpha^l_n &= \frac{{F^l_n}^2\left(t - 4i\right) t^3}{24 (1+it)}.\label{eq:sup_lim_alpha}
\end{align}
In the case of vanishing $\omega^l_n$ (e.g., a state $\ket{n}$ having no projection upon the blockaded states), one must use Eqs.~\eqref{eq:sup_lim_bark}–\eqref{eq:sup_lim_alpha} to define $\bar{K}^{ll}_n, \bar{b}^l_n, $ and $\alpha^l_n$. Note that Eq.~\eqref{eq:sup_psi_final} reduces to the solution of a freely diffusing atom in free space when $\omega^l_n$ and $F^l_n$ vanish; additionally, if $\omega^l_n$ vanishes but $F^l_n$ is non-zero, one finds that the squared wavefunction is centered at $s=F^l_n t^2/2$; i.e., that the center of the wavefunction follows the classical motion of an accelerated atom under a force $F^l_n$, as expected by Ehrenfest's theorem.

\subsection{Gram-Schmidt and disentangling transformations}

We detail here the Gram-Schmidt (GS) orthogonalization process which removes the independent degree of freedom $\Sigma=\sum_l x^l$ from the problem. We begin by considering the first spin pair separation in laboratory coordinates, which may be expressed as
\begin{equation}
    x^2 - x^1 = \left[x^1, x^2, x^3, \dots\right] \left[-1, 1, 0, \dots\right]^\text{T}.
\end{equation}
We construct the first GS unit vector
\begin{equation}
    \vec{e}_1 = \frac{1}{\sqrt{2}} \left[-1, 1, 0, \dots\right]^\text{T}.
\end{equation}
Note that we use lower indices to label unit vectors. Considering the second displacement vector $\vec{d}_2 = [0, -1, 1, 0, \dots]^\text{T}$, we next construct the second GS vector
\begin{equation}
    {\vec{e}_2}^{\:\prime} = \vec{d}_2 - \left(\vec{d}_2 \cdot \widehat{g}_1\right) \widehat{g}_1
\end{equation}
which normalizes to
\begin{equation}
    \vec{e}_2 = \frac{1}{\sqrt{2}} \left[-1, -1, 2, 0, \dots\right]^\text{T} / \sqrt{6}.
\end{equation}
The process continues until we generate the transformation $\bar{\mathbf{g}} = \bar{G} \mathbf{x}$ between laboratory coordinates $\mathbf{x}$ and GS coordinates $\bar{\mathbf{g}} \in \mathbb{R}^L$, where
\begin{equation}
    \bar{G} =
    \left[
        \begin{array}{cccc}
            \vec{g}_1 & \vec{g}_2 & \dots & \vec{g}_L
        \end{array}
    \right]^\text{T}
\end{equation}
such that
\begin{equation}
    \mathbf{x}^\text{T} M^\mathbb{R}_a \mathbf{x} = \bar{\mathbf{g}}^\text{T} \bar{G} M^\mathbb{R}_a \bar{G}^\text{T} \bar{\mathbf{g}}. \label{eq:m_gs_coords}
\end{equation}
When the GS process reaches the end of the spin chain, it will have only generated $L-1$ basis vectors; the additional independent degree of freedom $\bar{\mathbf{g}}^L = \sum_l x^l /\sqrt{L}$ will complete the GS basis and form the final row of $\bar{G}$, but it must not contribute to the sum in Eq.~\eqref{eq:m_gs_coords}. We may therefore discard the lowest row of $\bar{G}$, defining
\begin{equation}
    G = \left[
        \begin{array}{cccc}
            \vec{g}_1 & \vec{g}_2 & \dots & \vec{g}_{L-1}
        \end{array}
    \right]^\text{T}
\end{equation}
such that
\begin{equation}
    \mathbf{x}^\text{T} M^\mathbb{R}_n \mathbf{x} = \mathbf{g}^\text{T} S_n \mathbf{g} \label{eq:m_gs_coords_2}
\end{equation}
for $\mathbf{g} \in \mathbb{R}^{L-1}$ and where $S_n$ is the $(L-1){\times}(L-1)$ matrix
\begin{equation}
    S_n = G M^\mathbb{R}_n G^\text{T}.
\end{equation}
We note that the elements of $G$ are in fact given by the closed formula
\begin{equation}
    G^{lk} = \begin{cases}
        - 1 / \sqrt{l+l^2}, & k < l + 1 \\
        l / \sqrt{l+l^2}, & k=l+1 \\
        0, & k > l+1.
    \end{cases} \label{eq:gs_formula}
\end{equation}
Eq.~\eqref{eq:gs_formula} may be used instead of performing the full GS algorithm, though in practice the GS execution time is negligible.

Each $S_n$ generated in this manner is real and symmetric, such that it possesses an orthogonal transformation
$S_n = Q^\text{T}_n D_n Q_n$. Assembling this factorization with Eq.~\eqref{eq:m_gs_coords}, we have that
\begin{equation}
    \mathbf{x}^\text{T} M^\mathbb{R}_n \mathbf{x} = \mathbf{g}^\text{T} Q^\text{T}_n D_n Q_n \mathbf{g},
\end{equation}
from which we identify the $n$-disentangled coordinates
\begin{equation}
    \mathbf{s}_n = Q_n \mathbf{g} = Q_n G \mathbf{x},
\end{equation}
or $\mathbf{s}_n = T_n \mathbf{x}$ for $T_n = Q_n G \in \mathbb{R}^{L-1}{\times}\mathbb{R}^L$. Recall from the main text that we denote by $d^l_n$ the $l$th of $L-1$ total diagonal elements of $D_n$.

\subsection{Proof that $K_{nm}$ is invertible}

To prove that $K_{nm}$ is invertible, note first that Eq.~\eqref{eq:overlap} must converge to a finite scalar since it is the partial trace over a unitary propagation $\widehat{\mathcal{U}}_\text{tot}$ of an initially normalized state. Additionally, all $\left|C_n\right|>0$ and all $b_{nm}$ are finite by inspection of their definitions.

Suppose for contradiction that $K_{nm}$ is not invertible, such that it possesses a non-trivial null space $\mathcal{K}$. We may now make a change of integration variable in Eq.~\eqref{eq:overlap} to a basis of $\mathcal{K}$, such that the integration over any single coordinate $y$ in said basis has integrand $e^{-y 0 y + c}$ for finite $c$. This integral diverges over all space, and since $\left|C_n\right|>0$, so too must Eq.~\eqref{eq:overlap} diverge. This contradicts our opening remark; $K_{nm}$ must therefore be invertible.

\end{document}